# Enhanced controllable triplet proximity effect in superconducting spin-orbit coupled spin valves with modified superconductor/ferromagnet interfaces

A. T. Bregazzi[1a], J. A. Ouassou[2], A. G. T. Coveney[1], N. A. Stelmashenko[3], A. Child[1], A. T. N'Diaye[4], J. W. A. Robinson[3], F. K. Dejene[1], J. Linder[2] and N. Banerjee[5,1,b]

[1] Department of Physics, Loughborough University, Loughborough, LE11 3TU, United Kingdom

[2] Center for Quantum Spintronics, Norwegian University of Science and Technology, Norway

[3] Department of Materials Science & Metallurgy, University of Cambridge, 27 Charles Babbage Road, Cambridge CB3 0FS, United Kingdom

[4] Advanced Light Source, Lawrence Berkeley National Laboratory, Berkeley, California 94720, USA

[5] Department of Physics, Blackett Laboratory, Imperial College London, London SW7 2AZ, United Kingdom

Authors to whom correspondence should be addressed: [a)]A.T.Bregazzi@lboro.ac.uk, [b)]n.banerjee@imperial.ac.uk

In a superconductor/ferromagnet hybrid, a magnetically-controlled singlet-to-triplet Cooper pair conversion can modulate the superconducting critical temperature. In these triplet superconducting spin valves, such control usually requires inhomogeneous magnetism. However, in the presence of spin-orbit coupling from an interfacial heavy metal layer, the singlet/triplet conversion rate and thus critical temperature can be controlled via the magnetization direction of a single homogeneous ferromagnet. Here, we report significantly enhanced controllable pair conversion to a triplet state in Nb/Pt/Co/Pt superconducting spin valve in which Pt/Co/Pt is homogeneously magnetized and proximity-coupled to a superconducting layer of Nb. The Co/Pt interface furthest away from Nb is modified by a sub-nanometer-thick layer of Cu or Au. We argue that the enhancement is most likely associated from an improvement of the Co/Pt interface due to the insertion of Cu and Au layers. Additionally, the higher normalized orbital moments in Au measured using X-ray magnetic circular dichroism shows that increasing spin-orbit coupling enhances the triplet proximity effect – an observation supported by our theoretical calculations. Our results provide a pathway to enhancing triplet pair creation by interface engineering for device development in superspintronics.

Conventional superconductivity arises due to the formation of Cooper pairs between electrons with opposite spins. In a superconductor/ferromagnet (S/F) bilayer, such pairs can leak from S into F, causing the so-called proximity effect. Due to the strong exchange field in the F layers, this proximity effect is short-ranged. Microscopically, the two electrons of the Cooper pair occupy two different spin bands of the spin-split bands of the F layer resulting in a finite centre-of-mass momentum [1]. This results in an oscillatory dependence of the critical temperature ($T_C$) of the superconductor versus F layer thickness in addition to the strong $T_C$-suppression [2]. Moreover, in a superconducting spin valve (SSV) consisting of F/S/F layers, the $T_C$ can be controlled by modulating the relative magnetic moments of the F layers. In the parallel (anti-parallel) state, the higher (lower) net exchange field experienced by the Cooper pairs in S results in a lower (higher) $T_C$ as expected by theory [3–6] and generally seen experimentally [3,7,8] with some exceptions [9].

However, in a S/F/F or a F/S/F trilayer the $T_C$ of the S layer is lowered for an orthogonal magnetization alignment of the two F layers due to the formation of spin-aligned triplet Cooper pairs generated from the non-collinear alignment of the F layer moments. These triplets belong to the same spin-band and are less sensitive to the pair-breaking effects of the exchange field resulting in an enhanced proximity effect i.e. triplet pairs can pass much further into F than singlet pairs. This enhancement *spreads* the superconductivity in a larger volume allowing a greater modulation of $T_C$ by controlling the relative F layer moments [10–13].

Recent theoretical and experimental work [14-16] have demonstrated that in the presence of spin-orbit coupling (SOC), a triplet SSV can be achieved by controlling the magnetization direction of a single homogeneous F layer. This effect arises from an interplay of the SOC and the magnetic exchange field ($h_{ex}$) of the F layer [16-19]. Experimentally, this was demonstrated in Nb/Pt/Co/Pt where the Pt/Co/Pt trilayer has a weak Rashba SOC [15] due to the structural inversion asymmetry arising from the differences in microscopic nature of Pt/Co and Co/Pt interfaces [15,16,20].

Here, the $T_C$ modulation mechanism is different from the S/F/F trilayer and results from the selective opening and closing of the triplet channel depending on the angle of the magnetic moment with the film plane ($\theta$, Fig. 1a) of the single SOC F layer. This is understood by solving the Usadel equations with a SOC term. The SOC has a depairing effect on the triplets whose magnitude depends on the angle of magnetization through the relation: $E_t(\theta) = \epsilon + iD\alpha^2 (3 - \cos 2\theta)$, where $\epsilon$ is the quasi-particle energy and $D$ is the diffusion coefficient [15]. The magnitude of $E_t(\theta)$ is related to the depairing that triplets experience in the presence of SOC [16]. The magnitude of the SOC-part of the depairing term is $2D\alpha^2$ for in-plane (IP) magnetic alignment and $4D\alpha^2$ for out-of-plane (OOP) magnetic alignment (orientation shown in Fig. 1a). The formation of triplet pairs via an OOP alignment becomes energetically unfavorable, reducing the "leakage" of singlet pairs and enhancing the $T_C$ of the S layer. This has two important consequences. Firstly, the SOC-induced depairing makes the zero-field $T_C(H = 0, \alpha)$ for an S/F structure is higher for stronger Rashba coefficients α. Secondly, the larger depairing term for the OOP configuration compared to IP configuration allows an S/F bilayer with SOC to act as a triplet SSV. Experimentally we observe both these signatures here.

Although the SOC driven SSV effect is well established [15,21], there are important open questions. For example, how does the conversion efficiency depend on the SOC strength and what parameters can be used to tune this effect? The SOC strength dependence is not only interesting in the context of triplets where too high or weak SOC is predicted to lower the magnetic control over the singlet/triplet conversion efficiency [15], but experiments in 2DEGs have shown that modulating SOC with gate biasing can drive the superconductivity in to a topological regime [22]. In our fully metallic system gate biasing is not possible and SOC strength can be modulated by either changing the material or by modifying the interface. Following standard techniques from spintronics (23,24), here we modify the top Co/Pt interface in Nb/Pt/Co/Pt by inserting sub-nanometer Cu and Au layers. Cu is a lighter element improving interfacial spin transmission with Co [25] and heavier Au could enhance SOC [26] while reducing alloying effects with Co and Pt [27,28]. Previous studies in Pt/Co systems have shown sub-nanometer dusting improving the interface [29] and here we demonstrate a significant improvement of the magnitude and reproducibility of the SSV effect in these dusted samples compared to our previous results [15].

Thin-films were deposited by DC magnetron sputtering on unheated MgO (001) substrates placed on a rotating table in an ultrahigh vacuum chamber with the rotation speed controlling film thicknesses for a fixed magnetron power. Liquid-nitrogen was introduced in an outer jacket to achieve a base pressure below $3 \times 10^{-7}$ Pa. Deposition rates and nominal film thicknesses were estimated from control samples deposited onto a pre-fabricated substrate with a photolithographically defined polymer mask which was removed post-deposition using a lift-off process in acetone. The step-height of the thin film was measured using an atomic force microscope. Films were deposited at 1.5 Pa Ar pressure except Au which was deposited at 3.4 Pa. Nb was deposited at 60 W while the rest of the layers were deposited at 25 W. Figure 1a shows the stack with the applied field directions with IP (OOP) denoted by 0° (90°).

Four-point resistance measurements were performed on unpatterned spin-valves in the 2 – 8 K range. The $T_C$ was defined as the temperature corresponding to 50% of the resistive transition. Figure 1b shows a series of resistance (R) vs temperature (T) plots for a Nb(20)/Pt(2)/Co(1)/Au(0.6)/Pt(1.5) spin-valve versus applied OOP magnetic field. The numbers in parentheses denote thicknesses in nanometers. The black, red and blue curves correspond to resistive transitions recorded at 0, 100 and 500 mT OOP fields respectively, showing a progressive suppression of $T_C$ arising from the orbital depairing expected for a superconductor in an OOP field.

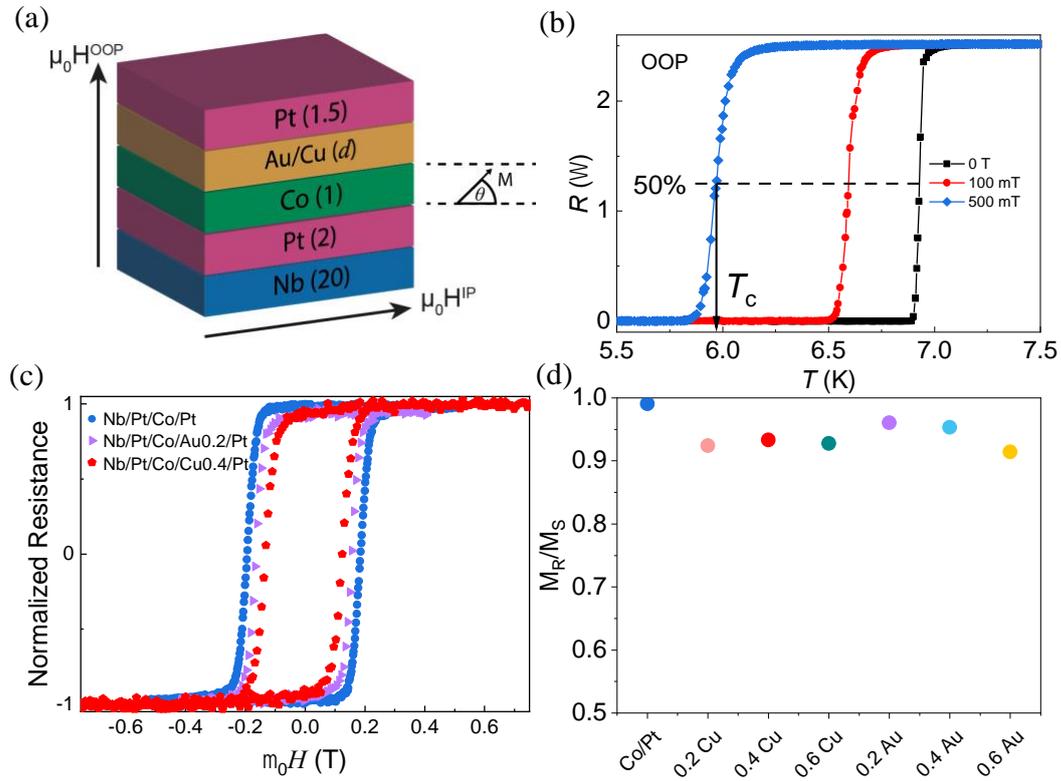

FIG 1. (a) A representative Nb(20)/Pt(2)/Co(1)/X(*d*)/Pt(1.5) spin-valve where X is Cu or Au and *d* = 0.2, 0.4 and 0.6 nm. The IP and OOP magnetic fields are indicated. (b) Resistance $R$ versus temperature $T$ for Nb(20)/Pt(2)/Co(1)/Au(0.6)/Pt(1.5) for OOP fields of 0 (black), 100 (red) and 500 (blue) mT respectively with the superconducting transition temperature, $T_C$, indicated. All lines are guides to the eye. (c) The AHE signal at 10 K. The legend shows the sample structures. (d) The ratio of remanent and saturation magnetization for all the samples.

Figure 1c shows the normalized anomalous Hall effect (AHE) at 10 K (above the $T_C$) for a Nb/Pt/Co/Pt (control) sample and two dusted samples where the top Co/Pt interface is modified by inserting a discontinuous layer of 0.2 nm or 0.4 nm Au or Cu respectively. Aside the differences in coercive fields, the shape of the curves is similar meaning that the dusted layers on top of the Co does not change its OOP anisotropy. Figure 1d shows that the ratio of the remnant to the saturation magnetization for all the samples has less than 10% variation indicating these are magnetically similar.

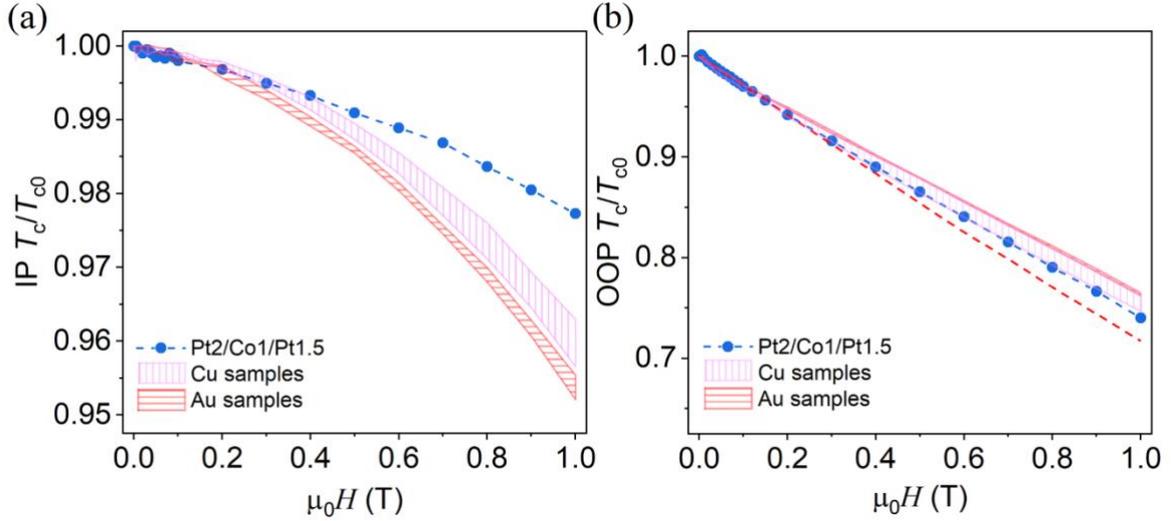

FIG 2. $T_C$ versus IP (a) and OOP (b) applied fields for Nb(20)/Pt(2)/Co(1)/Pt(1.5) (blue data points), Nb(20)/Pt(2)/Co(1)/Cu($d$)/Pt(1.5) (purple shaded region) and Nb(20)/Pt(2)/Co(1)/Au($d$)/Pt(1.5) (red shaded region) with $d$ = 0.2, 0.4 and 0.6 nm. The data for Au $d$ = 0.4 nm is shown separately as a red dashed line (see text). $T_C$ at each field is divided by its corresponding $T_C$ at zero field denoted by $T_{C0}$. The dotted lines and shaded regions are for guide to the eyes.

Figure 2 shows $T_C(H)$ for the control and dusted samples (averaged for several measurements with a spread of less than 5 mK) normalized by the zero-field $T_C$ values ($T_{C0}$) of each sample. Instead of plotting the $T_C$ difference with the control samples, the normalized plots allow us to directly compare the $T_C(H)$ between different samples and removes any effect of $T_C$ suppression which might arise due to small S layer thickness variations between different runs. Figure 2a shows the normalized $T_C(H)$ for IP fields for the control Nb/Pt/Co/Pt sample together with six Cu and Au dusted samples with dusting layer thicknesses of 0.2 nm, 0.4 nm and 0.6 nm. We did not see any specific trend of $T_C(H)$ as a function of the dusted layer thicknesses apart from an overall larger reduction in $T_C$ for samples with thicker dusted layers and in the figure the entire range of $T_C$ variation is represented as a hatched region for clarity. For all the dusted samples the $T_C$ reduction is significantly larger (~3-4%) than the pristine Nb/Pt/Co/Pt sample (~2%). This corresponds to a $T_C$ suppression of 0.324 K (0.6 Au) to 0.240 K (0.2 Cu) for the dusted samples compared to 0.135 K for the pristine sample. These $T_C$ suppressions are much larger than a pure Nb and Nb/Co/Pt sample of similar thicknesses (0.115 K and 0.120 K respectively, not shown here) which can be explained by a weak field-induced depairing for superconducting films as we observed before [15]. All ten dusted samples measured for this study showed consistently large $T_C$ drops including samples with a 1.2-nm-thick layer of Co instead of 1 nm (not shown here). This is in contrast to Nb/Pt/Co/Pt where there were some variation in the magnitude of the $T_C$ drop between different deposition runs.

Figure 2b shows similar $T_C(H)$ for OOP fields. As expected, there is a strong $T_C$ suppression arising from orbital depairing which can be understood from the Ginzburg-Landau equations for a superconducting thin film in a perpendicular field. The reduction in $T_c$ ($= \Delta T_C$) is directly proportional to the OOP field (Ref 15, equation 9) and for our control samples of pure Nb or Nb/Co bilayer, this drop is ~35% in the same field range. Our previous studies have shown that for Nb/Pt/Co/Pt samples this drop is significantly lower owing to triplet generation as explained earlier, but here we see that the dusted samples have an even smaller $T_C$ drop. The only outlier is the sample with 0.4 nm Au thickness for OOP fields which is shown as a separate dotted line.

The key feature in these resistance measurements is the significantly larger $T_C$ drop for IP fields compared to those due to a weak field-dependent depairing in pure Nb or Nb/Co bilayer. Similarly, for OOP fields, the $T_C$ drops in samples containing Pt at the interface between Nb and Co is significantly smaller than is expected from orbital depairing. These observations are consistent with singlet-triplet conversion mediated by SOC [15]. We now focus on the enhanced $T_C$ drop in all the dusted samples compared to pristine Nb/Pt/Co/Pt. As mentioned earlier, Cu improve spin transmission to Co [25] and most likely, in this case, provides a buffer layer over Co preventing interdiffusion. Au and Co are immiscible, possibly leading to a lower alloying between Co and Pt. Furthermore, Au/Co multilayers are known to have high SOC [30], which, with an improved interface, can lead to an increased (reduced) drop in the $T_C$ for IP (OOP) fields.

We note a second feature in Fig. 2: the $T_C$ drop for IP fields gets larger at higher fields. If the $T_C$ drop was purely controlled by the orientation of the magnetic moment, the effect should disappear beyond the saturation field of the Co layer (~120 – 150 mT). Similarly, for OOP fields, the $T_C$ *recovery* effect should stop beyond this saturation field which does not happen and the slope of the $T_C(H)$ line is constant even at higher fields in Fig. 2b. We currently do not understand this but we note that similar effects have been noted before in SSV where $T_C$ drop continued beyond saturation of the F layer [12]. This was attributed to uncompensated spins at the interface with much higher anisotropy than the bulk of the F layer. A similar situation could explain our results, particularly when uncompensated spins have been reported at the surface of ultra-thin Co layers [12,31,32] although alternative explanations cannot be ruled out.

To estimate the SOC, we performed X-ray Magnetic Circular Dichroism (XMCD) measurements at room temperature at the magnetic spectroscopy beamline (4.6) at the Advanced Light Source, Lawrence Berkeley National Laboratories. The X-ray beam, incident perpendicular to the sample surface, was scanned across the Co $L_{2/3}$ edge by varying the energy and the X-ray absorption spectra (XAS) recorded using the total electron yield method. A magnetic field of 1.8 T was applied normal to the sample surface to saturate the Co layer when recording the XAS. The XMCD is the difference of the two XAS spectra recorded at positive and negative magnetic fields corresponding to two helicities. We recorded 10 XAS scans for each sample which was averaged to get a single XAS spectrum. Applying the sum rules [33] we extracted the normalized orbital to spin moment ratio ($m_{\text{orb}} / m_{\text{spin}}$) which is proportional to the SOC [34].

From Fig. 3, we see that the normalized orbital moment lies between 0.12 to 0.14 for Pt/Co/Pt and Cu dusted Pt/Co/Pt samples albeit with slightly lower ratios for Cu dusted samples which agrees with literature values [35]. In contrast, for Au dusted samples, we observe a steady increase in this ratio with increasing thickness of the Au dusted layer which reaches a maximum of ~0.28 which is twice the value for the pristine and Cu dusted samples. This strong enhancement in SOC possibly arises from using Au which in general shows a higher SOC than using Pt, and coupled with the improved interfaces, can lead to significantly higher SOC. On the other hand, Cu being a lighter element reduces the SOC by increasing the distance between the two interfaces.

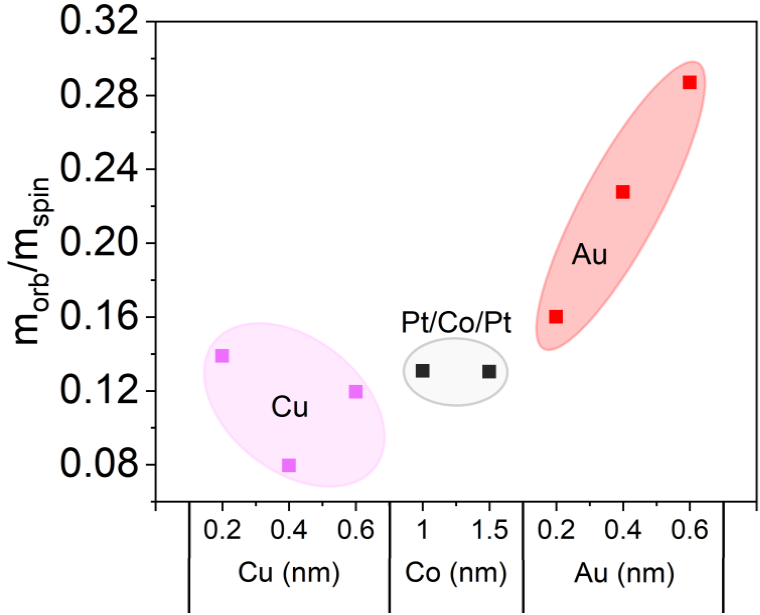

Figure 3. Orbital to spin moment ratio calculated at room temperature for Nb(20)/Pt(2)/Co(1)/X(d)/Pt(1.5). Here, X is Cu or Au and denotes the dusted layer at the top Co/Pt interface with thickness $d$ in nanometers. This thickness is plotted along the X axis. In the middle section, the thickness denoted is for the Co layer of the pristine samples. The shaded regions are guide to the eye.

Although it is difficult to deconvolute the enhanced control and magnitude of the triplet proximity effect due to increased SOC strength and improved interfaces, we can nonetheless perform a semi-quantitative estimation as explained below.

We performed numerical calculations using the Usadel equation for diffusive superconductivity [16,36] with a bisection algorithm for determining $T_C$ [37]. The Nb/Pt/Co/X/Pt junctions were approximated as effective S/F bilayers. Thin individual layers in the effective F (Pt/Co/X/Pt) are best described using ballistic models. However, multiple layers and interfaces enhances scattering, and the effective F can be approximated as a diffusive metal. We model this region as having a homogeneous ferromagnetic exchange field oriented partly OOP in the absence of applied fields, rotating further IP or OOP depending on the direction and magnitude of the applied field. We describe the Rashba SOC in F by the dimensionless parameter $\alpha\xi$, where $\xi$ is the superconducting coherence length and the magnitude of $\alpha$ depends on the choice of dusting. For OOP fields, a linear-in-$H$ orbital depairing effect is included. Except for specific field values and SOC magnitude $\alpha$, the numerical model and parameters are the same as presented in Sec. IV of Ref. [15] albeit some expected quantitative differences from Ref. [15]. For instance, the exact dependence of magnetization on the applied field, or the interface transparencies may individually vary leading to only quantitative differences for calculated $T_C(H, \alpha)$, as detailed in Ref. [15]. The magnetization is modelled according to the equation $M = M_0 + \delta M \tanh(H/H_0)$ where $M$ is the magnetization and $H$ is the applied field similar to Ref. [15], and the parameters $\delta M$ and $H_0$ parametrize magnetization changes and saturation in an applied field. The numerics assumes that the magnetization points 45º out of the thin-film plane in the absence of $H$. When an IP field is applied, $M$ is rotated 30º more into the plane; when an OOP field $H$ is applied, it is instead rotated 30º more out of the plane. This corresponds to the average case considered in Ref. [15], where we studied the $T_C$ variation with all these parameters. We refer the reader to Ref. [15] for a detailed discussion.

Figure 4 shows the dependence of $T_C$ of the S/F bilayer versus the effective Rashba coefficient $\alpha$ in the F layer. Firstly, the numerical results show that $T_C(H,\alpha)$ increases as a function of $\alpha$ for *all* applied fields $H$. Secondly, the $T_C$ increases more when OOP fields are applied cf. IP fields—and the magnitude of this triplet SSV effect also increases as a function of the Rashba coefficient $\alpha$. This is consistent with the fact that changing the strength of SOC can tune the triplet proximity effect and qualitatively in agreement with our experimental findings and explanation based on triplet depairing. We also observe that the zero-field $T_C$ of the dusted samples (except for Cu or Au ($d$=0.4 nm)) are approximately 300-500 mK higher than the Nb/Pt/Co/Pt sample. This is consistent with our model where the zero-field $T_C$ is higher for S/F hybrids with larger SOC (see Introduction). This does not explain higher $T_C$ for Cu dusted samples where the SOC is similar to undusted samples possibly because our simplified model does not account for variation in effective diffusion constant and tunneling conductances in dusted samples. Furthermore, effects like decrease in $T_C$ due to enhanced boundary transparency from dusting is not accounted for in our model.

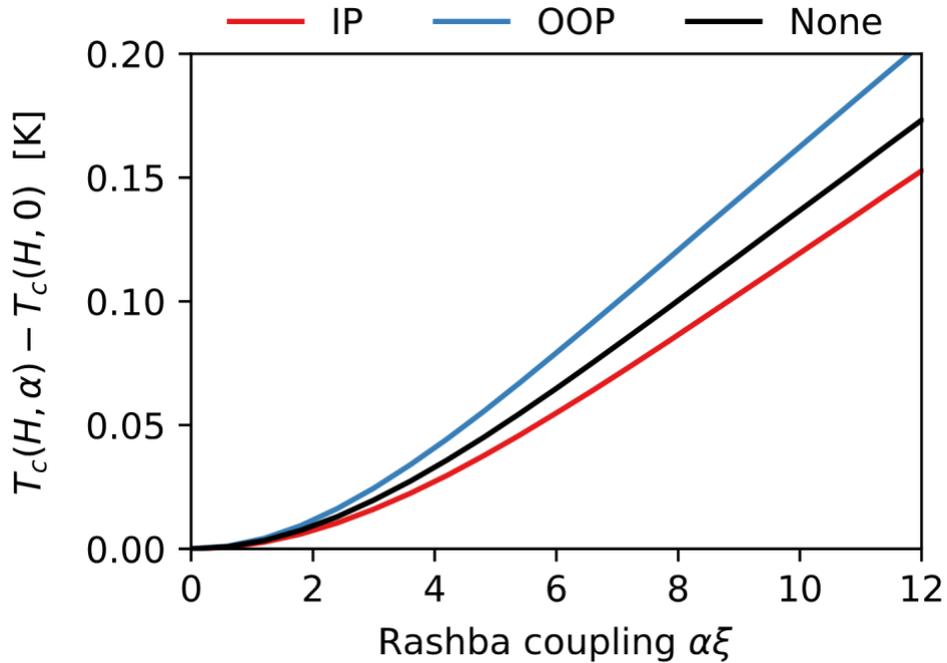

FIG 4. Numerical results for the $T_C(H,\alpha)$ of an S/F bilayer as a function of the Rashba coefficient $\alpha$. The three curves correspond to (i) an IP field $\mu_0 H = 100$ mT, (ii) an OOP field $\mu_0 H = 100$ mT, (iii) zero field. To highlight the difference attributable to SOC (as opposed to e.g. orbital depairing effects), we have subtracted $T_C(H,0)$ of a corresponding S/F structure with no SOC under the same applied field.

In summary, we have demonstrated that SOC-driven triplet proximity effects and its magnetic control are strongly enhanced by engineering heavy-metal/ferromagnet interfaces. This opens up opportunities to design functional devices – for example, in SOC-driven spin-polarized Josephson junctions where arbitrary phase bias can be achieved with possible applications in superconducting quantum circuits [38].

**Acknowledgements:**

ATB was supported by Loughborough University. NB acknowledges funding from the EPSRC (EP/S016430/1) and would like to thank Laura Stuffins for assistance on some initial part of the project. JL and JAO are supported by the Research Council of Norway through Grant No. 323766 and its Centres of Excellence funding scheme Grant No. 262633 "QuSpin." Support from Sigma2 - the National Infrastructure for High Performance Computing and Data Storage in Norway, project NN9577K, is acknowledged. J.W.A.R. acknowledge the EPSRC through the Core-to-Core International Network "Oxide Superspin" (EP/P026311/1).